\documentclass[12pt,preprint]{aastex}

\def\be{\begin{equation}}
\def\ee{\end{equation}}
\def\b{_{\rm b}}
\def\bh{_\bullet}
\def\BBH{_{\rm BBH}}
\def\c{_{\rm c}}
\def\coll{_{\rm coll}}
\def\diff{^{\rm diff}}
\def\D{_{\rm D}}
\def\e{_{\rm e}}
\def\ej{_{\rm ej}}

\def\full{^{\rm full}}
\def\h{_{\rm h}}
\def\ij{_{ij}}
\def\lc{_{\rm lc}}

\def\sgr{{\rm Sgr\,A^*}}

\newcommand{\half}{{\textstyle{1\over2}}}

\def\d{{\rm d}}

\def\bfx{{\bf x}}
\def\bfv{{\bf v}}
\def\bfr{{\bf r}}
\def\bfe{{\bf e}}

\def\kms{{\rm\,km\,s^{-1}}}
\def\kpc{{\rm\,kpc}}
\def\msun{{\rm\,M_\odot}}
\def\rsun{{\rm\,R_\odot}}

\def\max{{\rm\,max}}
\def\mas{{\rm\,mas}}
\def\pc{{\rm\,pc}}
\def\mpc{{\rm\,mpc}}
\def\yr{{\rm\,yr}}

\def\au{{\rm\,AU}}

\begin{document}

\title{Ejection of hypervelocity stars by the
(binary) black hole(s) in the Galactic center}
\author{Qingjuan Yu$^1$ and Scott Tremaine$^2$}
\affil{$^1$Canadian Institute for Theoretical Astrophysics, 60 St.\ George
Street, Toronto, Ontario M5S 3H8, Canada \\ $^2$Princeton University
Observatory, Peyton Hall, Princeton, NJ~08544-1001, USA}

\begin{abstract}
\noindent
We study three processes that eject hypervelocity ($>10^3\kms$) stars from the
Galactic center: (i) close encounters of two single stars; (ii) tidal breakup
of binary stars by the central black hole, as originally proposed by Hills;
and (iii) three-body interactions between a star and a binary black hole
(BBH).  Mechanism (i) expels hypervelocity stars to the solar radius at a
negligible rate, $\sim 10^{-11}\yr^{-1}$.  Mechanism (ii) expels hypervelocity
stars at a rate $\sim 10^{-5}(\eta/0.1)\yr^{-1}$, where $\eta$ is the fraction
of stars in binaries with semimajor axis $a\b\la 0.3\au$.  For solar-mass
stars, the corresponding number of hypervelocity stars within the solar radius
$R_0=8\kpc$ is $\sim 60(\eta/0.1)(a\b/0.1\au)^{1/2}$. For mechanism (iii),
$\sgr$ is assumed to be one component of a BBH.  We constrain the allowed
parameter space (semimajor axis, mass ratio) of the BBH.  In the allowed
region (for example, semimajor axis of $0.5\times 10^{-3}\pc$ and mass ratio
of 0.01), the
rate of ejecting hypervelocity stars can be as large as $\sim 10^{-4}\yr^{-1}$
and the expected number of hypervelocity stars within the solar radius can be
as large as $\sim 10^3$. Hypervelocity stars may be detectable by the next
generation of large-scale optical surveys.
\end{abstract}


\section{Introduction}\label{sec:intro}

\noindent
The Milky Way, as well as most nearby elliptical galaxies and disk galaxies
with spheroids, is believed to house a massive black hole (BH) at its center
(e.g.\
\citealt{Schodel02,Genzel03,Schodel03,eis03,Ghez03a,Ghez03b,Ghez03c,Gebhardt03}).
The deep gravitational potential well close to the BH enables relativistic
phenomena, such as the relativistic motion of jets associated with some active
galactic nuclei (e.g.\ \citealt{BBR84}). \citet{H88} first pointed out that
stars may also be expelled from the vicinity of the BH with high velocities.
The main features of Hills' argument can be explained simply. Consider a star
with specific energy $E\equiv{1\over2} v^2+\Phi(r)$, where $v$ and $\Phi(r)$
are the velocity and specific potential of the star at position $\bfr$. Close
to the BH, $\Phi(r)\simeq-GM\bh/r$ where $G$ is the gravitational constant and
$M\bh$ is the BH mass.  Now suppose that the star approaches close enough to
the BH so that $|\Phi(r)|\gg|E|$. In this case, the star will have a velocity
$v=\sqrt{2[E-\Phi(r)]}\simeq\sqrt{2GM\bh/r}=2.9\times 10^3\kms
(M\bh/10^6\msun)^{1/2}(1 \mpc/r)^{1/2}$, where $\mpc\equiv 10^{-3}\pc$ will be
often used as a length unit in this paper.
If the star then suffers a velocity
change $\delta v\ll v$ (e.g.\ due to interactions with surrounding stars or
the tidal disruption of a close binary) and the increase of the specific
energy of the star $\delta E={1\over2}(v+\delta v)^2-{1\over2}v^2\simeq
v\delta v$ is much larger than $|E|$, the star will escape from the BH with
velocity roughly given by
\be
\sqrt{2v\delta v}\simeq 1.5\times 10^3\kms(M\bh/10^6\msun)^{1/4}
(1 \mpc/r)^{1/4}(\delta v/400\kms)^{1/2}.
\label{eq:vdeltav}
\ee
The discovery of such hypervelocity stars---stars with velocity exceeding
$10^3\kms$---would provide strong evidence for the existence of a massive
central BH in our Galaxy---if any is still needed---based solely on stellar
kinematics in the {\it outer} parts of the Galaxy.  The properties of
hypervelocity stars would also illuminate the nature of the innermost regions
of galaxies, such as the stellar kinematics, the age and metallicity
distribution, the number and mass of the BHs, etc.

In this paper, we will study the expected properties of such hypervelocity
stars. We consider three possible scattering processes.

\begin{enumerate}

\item Gravitational encounters of single stars. The gravitational
interaction of two stars of masses $m_i$ and $m_j$ leads to a velocity change
of star $m_i$,
\begin{eqnarray}
\delta v
& = & \frac{2Gm_j}{[G^2(m_i+m_j)^2/w^2\ij+Y\ij^2w^2\ij]^{1/2}} \nonumber \\ &
\le & \frac{2Gm_j}{[2G(m_i+m_j)Y\ij]^{1/2}} \nonumber \\ & = & 4.4\times
10^2\kms \left(\frac{2m_j}{m_i+m_j}\right)^{1/2}
\left(\frac{m_j}{1\msun}\right)^{1/2}\left(\frac{1\rsun}{Y\ij}\right)^{1/2}
\label{eq:deltav}
\end{eqnarray}
where $Y\ij$ is the impact parameter and $w\ij$ is the initial relative
velocity of the two stars \citep{BT87}. Only rare close encounters (impact
parameters of a few solar radii) can produce velocity changes large enough for
a star to escape from the BH at high velocity.

\item Tidal breakup of a binary star.  We denote the component masses of the
binary as $m_1$ and $m_2$.  A binary with semimajor axis $a\b$ will be tidally
disrupted by the massive BH when it passes closer to the BH than the Roche or
tidal radius $\sim a\b[M\bh/(m_1+m_2)]^{1/3}\simeq
10\au (a\b/0.1\au)[M\bh/10^6(m_1+m_2)]^{1/3}$.  During the tidal breakup, each
star will receive a velocity change of order its orbital velocity relative to
the center of mass of the binary star, that is, the velocity change of $m_1$
is about \citep{H88,GQ03}
\begin{eqnarray}
\delta v & \sim &
\sqrt{\frac{G(m_1+m_2)}{a\b}}\left(\frac{m_2}{m_1+m_2}\right) \nonumber \\
 & \simeq & 67\kms \left(\frac{2m_2}{m_1+m_2}\right)^{1/2}
\left(\frac{m_2}{1\msun}\right)^{1/2}\left(\frac{0.1\au}{a\b}\right)^{1/2}.
\label{eq:deltavb}
\end{eqnarray}
If one star gains energy during this process, the other will lose energy and
become more tightly bound to the BH.  Comparing equations (\ref{eq:deltav})
and (\ref{eq:deltavb}), we see that the semimajor axis of the binary star
$a\b$ in equation (\ref{eq:deltavb}) plays the same role as the impact
parameter $Y\ij$ in equation (\ref{eq:deltav}). 

\item Ejection by a binary black hole (BBH). This is actually a special case
of mechanism (1), in which the perturbing star $m_j$ is one component of a
BBH.  The existence of BBHs in some galactic centers is a necessary
consequence of hierarchical galaxy formation
\citep{BBR80,Y02,mm02,vhm02,kom03}\footnote{We do not consider the possibility
of three or more massive BHs in galactic centers, although this configuration
may also occur in some galaxies.}.  We denote the component masses of the BBH
as $M_1$ and $M_2$ ($M_1\ge M_2$).  If the semimajor axis $a\bh$ of the BBH is
smaller than a certain value $a\h$($\equiv GM_2/4\sigma\c^2$, where $\sigma\c$
is the one-dimensional velocity dispersion of the galactic center, see
\citealt{Q96}, eq.\ 29), most of the low angular momentum stars that pass
through its vicinity will be expelled with an energy gain after one or several
close encounters with the BBH, and the average velocity increase of these
stars is given by:
\be
\delta v\sim F\delta t \sim 1.5\times 10^3 \kms
\left(\frac{2M_2}{M_1+M_2}\right)^{1/2}
\left(\frac{M_2}{10^6\msun}\right)^{1/2}
\left(\frac{1 \mpc}{a\bh}\right)^{1/2},
\label{eq:deltavbbh}
\ee
where the force per unit mass from the BH $M_2$ is $F\sim GM_2/a^2\bh$ and the
interaction time $\delta t\sim [G(M_1+M_2)/a^3\bh]^{-1/2}$ \citep{Q96}. Note
that $\delta v$ in equation (\ref{eq:deltavbbh}) is independent of the stellar
mass, while $\delta v$ in equations (\ref{eq:deltav}) and (\ref{eq:deltavb})
depends on the mass of the ejected star; otherwise the three equations have
similar form.

\end{enumerate}

This paper is organized as follows.  In \S\ref{sec:Milky}, we review the data
on the inner structure of the Galaxy, such as the mass of the central BH,
stellar density, etc.  In \S\ref{sec:rates}, we study the rates of ejecting
hypervelocity stars due to the three processes discussed above.  A discussion
and conclusions are given in \S\ref{sec:discon}.

\section{The Galactic center}\label{sec:Milky}

\noindent
Measurements of stellar radial velocities and proper motions at radii
$r\ga 10\mpc$, the discovery of variable X-ray emission around
Sagittarius A$^*$, and the tracing of the Keplerian orbits of stars at radii
$r\sim 0.1$--$1\mpc$ have provided strong evidence for the existence
of a massive BH in the center of the Galaxy
\citep{Genzel00,Baganoff01,Schodel02,Genzel03,Schodel03,eis03,Ghez03a,Ghez03b,Ghez03c}.
We expect that the distribution of old stars should be approximately spherical
and isotropic, since the two-body relaxation time is short, and this
expectation is consistent with the observations \citep{Genzel00}.  We may
describe the mass distribution near the center of the Galaxy by the
combination of a point mass of $3.5\times 10^6\msun$ (obtained by averaging
the values $(4.0\pm0.3)\times10^6\msun$ in \citealt{Ghez03c} and
$(2.9\pm0.2)\times10^6\msun$ in \citealt{Schodel03}, Fig. 11), plus the
visible stellar system with mass density
\be
\rho(r)=Ar^{-\alpha}\exp(-r^2/r\b^2).
\label{eq:rho}
\ee
Here $\alpha=1.8$, the characteristic radius of the bulge is $r\b=1.9\kpc$,
and the normalization $A$ can be obtained by setting
$\rho(r=0.16\pc)=3.6\times 10^6\msun\pc^{-3}$ \citep{DB98,Schodel03}. The
density distribution in equation (\ref{eq:rho}) has a cusp at small radii,
$\rho\propto r^{-\alpha}$ as $r\rightarrow 0$, consistent with theoretical
models of a stellar system with a central BH and all stars having the same
mass ($\alpha=1.75$, e.g.\
\citealt{BW76}). At the smallest radii, $r\la 10\arcsec=0.4\pc$, infrared
observations suggest that $\alpha$ decreases to 1.4 \citep{Genzel03}, somewhat
below the theoretical models.

Throughout this paper we shall assume that the distance to the Galactic center
is $R_0=8\kpc$ \citep{eis03}. If not otherwise stated, we assume for
simplicity that the stars in the Galactic center have the solar mass and
radius.  The effect of the generalization to a
distribution of masses and radii will be discussed briefly in
\S\ref{sec:discon}.

We define the distribution function (DF) $f(\bfx,\bfv)$ of a stellar system so
that $f(\bfx,\bfv)\,\d^3\bfx\d^3\bfv$ is the number of stars within a
phase-space volume $\d^3\bfx\d^3\bfv$ of $(\bfx,\bfv)$.  By Jeans's theorem,
the DF in a spherical and isotropic stellar system depends on $(\bfx,\bfv)$
only through the specific energy $E$.  The DF $f(E)$ in the Galactic center,
which will be used to calculate the ejection rates of hypervelocity stars
below, can be obtained from equation (\ref{eq:rho}) and the Eddington formula
(eq.~4-140a in \citealt{BT87}).  The potential in the stellar system
$\Phi(r)=\Phi\bh(r)+\Phi_*(r)$ is contributed by both the central BH
($\Phi\bh$) and the bulge stars ($\Phi_*$); however, at small radii $r\la
1\pc$ the gravitational potential contributed by stars can be ignored. In this
region we have $\Phi=\Phi\bh=-GM\bh/r$, $\rho(r)\propto r^{-\alpha}$, and the
stellar DF is
\be
f(E)=C|E|^{\alpha-3/2},
\label{eq:fE}
\ee
where
\be
C=2^{-3/2}\pi^{-3/2}\frac{\Gamma(1+\alpha)}{\Gamma(\alpha-{1\over2})}
\cdot\frac{\rho(r)}{m_*}\sigma(r)^{-2\alpha}, 
\qquad \sigma(r)\equiv\sqrt{\frac{GM\bh}{r}},
\label{eq:C}
\ee
and $m_*$ is the stellar mass. Note that $C$ is independent of radius since
the radial variations of $\rho(r)$ and $\sigma(r)$ cancel.

\section{Rates of ejecting stars with high velocities}\label{sec:rates}

\noindent
In this section, we study the rates of ejecting stars with high velocities
from the Galactic center.  We will consider in turn the three ejection
mechanisms described in \S\ref{sec:intro}.

\subsection{Close encounters between two single stars}
\label{sec:single}

\noindent
Two-body gravitational encounters between single stars can eject one of the
two stars \citep{H69,LT80}.  Consider a test star with mass $m_i$ and 
velocity $\bfv_i$ that encounters a star with velocity $\bfv_j$.  The test
star will experience a velocity change from $\bfv_i$ to $\bfv_i+\bfe\ij$ (see
eq.~\ref{eq:deltav} for $|\bfe\ij|$).  If the test star is moving through a
background of stars with DF $f(\bfr,\bfv)$, the probability
that this star will suffer an encounter with a background star during time $\d
t$, $\Gamma_i(\bfr,\bfv_i)\d t$, is given by:
\be
\Gamma_i(\bfr,\bfv_i)\d t=\d t\int\d^3\bfv_j~f(\bfr,\bfv_j)w\ij
\int\d Y\ij^2\int\d\Psi\ij,
\label{eq:Gammaia}
\ee
where $Y\ij$ is the impact parameter, $w\ij=|\bfv_i-\bfv_j|$ is the relative
velocity and $\Psi\ij$ is the angle between the $(\bfv_i,\bfv_j)$ plane
and the $(\bfv_i-\bfv_j,\bfe\ij)$ plane \citep{H60}.
If the velocity distribution of field stars is isotropic, we have
\be
\Gamma_i(\bfr,\bfv_i)\d t=\d t\int\d v_j~\frac{2\pi v_j}{v_i}f(\bfr,v_j)
\int\d w\ij~w^2\ij \int\d Y\ij^2\int\d\Psi\ij,
\label{eq:Gammaib}
\ee
where $v_j=|\bfv_j|$.  The probability that the test star is expelled 
with a velocity higher than $v_0$ at infinity can be obtained by
restricting this integral to the encounters that change the value of its
velocity from $v_i=|\bfv_i|$ to a value larger than $v\e$, i.e.,
\be
|\bfv_i+\bfe\ij|\ge v\e.
\label{eq:condition}
\ee
Here ${1\over2}v_0^2+\Phi(r=\infty)={1\over2}v\e^2+\Phi(r)\equiv E_0$, and we
shall assume that $\Phi(r=\infty)=0$.  We call $v_0$ the ejection speed and
$E_0$ the ejection energy.  Thus, the rate of ejecting stars with ejection
speed higher than $v_0$ due to two-body encounters in a unit volume around
position $\bfr$, ${\cal R}(\bfr,v_0)$, is given by:
\begin{eqnarray}
{\cal R}(\bfr,v_0) & = & \int\d^3\bfv_i~f(\bfr,\bfv_i)\Gamma_i(\bfr,\bfv_i)
\nonumber \\
& = & 8\pi^2\int\d v_i \int\d v_j \int\d w\ij
\int\d Y\ij^2\int\d\Psi\ij f(\bfr,v_i)f(\bfr,v_j)v_iv_j w\ij^2,
\label{eq:calR}
\end{eqnarray}
where the integration variables should satisfy the inequality
(\ref{eq:condition}).  The total rate of ejecting stars with ejection speed
higher than $v_0$ is given by
\be
\frac{\d N\ej(v_0)}{\d t}=4\pi\int_0^\infty {\cal R}(\bfr,v_0)r^2\d r.
\label{eq:dNej}
\ee

\begin{figure}
\epsscale{0.6}
\plotone{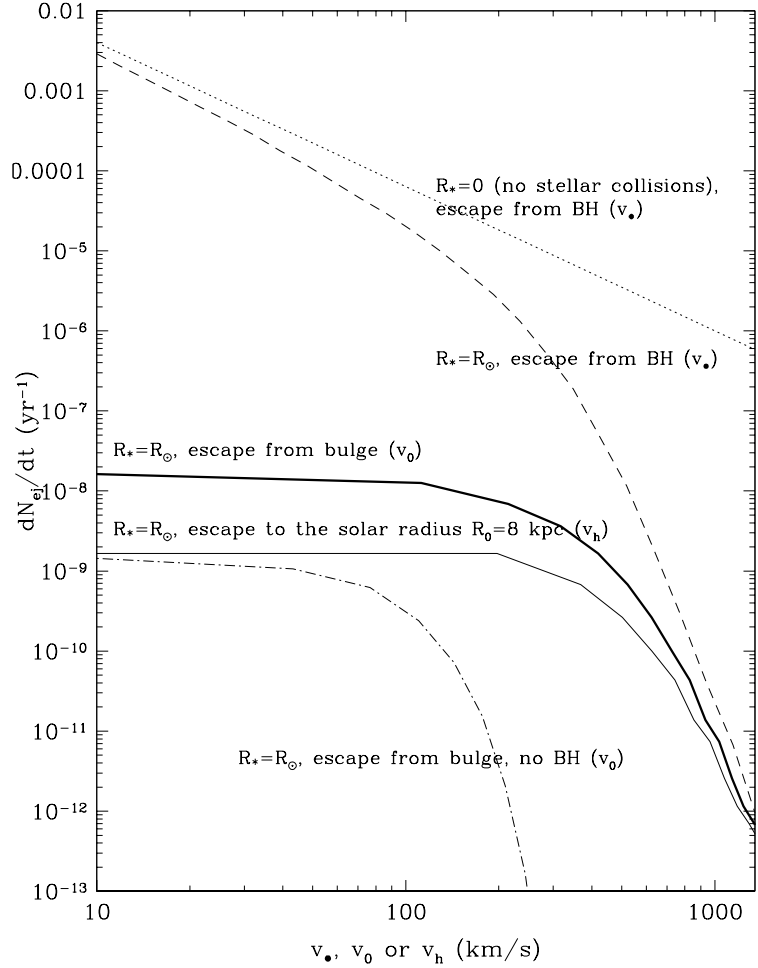}
\caption{Estimated rates of ejecting stars from the Galactic center by two-body
encounters of single stars. Here $v\bh$ is the velocity after escaping from
the BH, $v_0$ is the velocity after escaping from the bulge, and $v\h$ is the
velocity after reaching the solar radius.  The dotted line (a function of
$v\bh$) is obtained by treating stars as point masses with zero stellar radius
and ignoring the gravitational potential contributed by bulge stars (see
eq.~\ref{eq:dNejs}). The dashed line (a function of $v\bh$) is obtained by
assuming that all stars have the solar radius.  The bold solid line (a
function of $v_0$) is the ejection rate including the effects of both non-zero
stellar radii and the gravitational potential of the bulge. The thin solid
line (a function of $v\h$) is the rate at which ejected stars reach the solar
radius (see eqs.~\ref{eq:deltaPhi} and \ref{eq:vh}).  The current of
hypervelocity stars with velocity $>700\kms$ past the solar radius is
negligible (less than 1 per Hubble time). The dot-dashed line is the ejection
rate including the effects of both non-zero stellar radii and the
gravitational potential of the bulge, but assuming that there is no central
BH. The existence of the BH increases the ejection rate, especially at high
velocities. }
\label{fig:single}
\end{figure}

The ejection rate can be obtained analytically in some cases.  Ignoring
possible stellar collisions (when the impact parameter $Y\ij$ is small enough)
and treating stars as point masses with zero radius, equation (\ref{eq:calR})
can be simplified as follows \citep{H69,LT80}:
\be
{\cal R}(\bfr,v_0)=\frac{2^{13/2}}{3}\pi^3 G^2m_*^2
\int_\Phi^0\d E f(E)
\int_{E_0+\Phi-E}^0 \d E' f(E')\frac{(E'+E-E_0-\Phi)^{3/2}}{(E_0-E)^2}.
\label{eq:calRs}
\ee
By ignoring the gravitational potential contributed by stars (e.g.\ in the
region close to the BH) and applying equation (\ref{eq:fE}) in equations
(\ref{eq:dNej}) and (\ref{eq:calRs}), the total ejection rate of stars as a
function of ejection speed can be found to be \citep{LT80}:
\be
\frac{\d N\ej(v\bh)}{\d t} = 
\frac{2^{5/2}\pi^{3/2}}{3}\frac{G^5M\bh^3\rho^2}{\sigma^9}
\left(\frac{E\bh}{\sigma^2}\right)^{2\alpha-9/2}
\frac{\Gamma^2(1+\alpha)\Gamma({9\over2}-2\alpha)}
{(3-\alpha)(2-\alpha)}.
\label{eq:dNejs}
\ee
Although the ejection rate in this equation is integrated over radius, we have
expressed it in terms of the velocity dispersion $\sigma(r)$ and the density
$\rho(r)$, given by equation (\ref{eq:C})---these can be evaluated at any
radius since their radial dependences cancel out in equation
(\ref{eq:dNejs}). Note that in this formula the ejection speed $v\bh$
represents the velocity after escaping the gravity of the central BH but
before climbing through the bulge, and the ejection energy
$E\bh={1\over2}v\bh^2$.  Using $\alpha=1.8$, BH mass $M\bh=3.5\times
10^6\msun$, and the stellar mass density distribution $\rho(r)=Ar^{-\alpha}$
(see eq.~\ref{eq:rho} and the subsequent text) in equation (\ref{eq:dNejs}),
we obtain the ejection rate shown in Figure~\ref{fig:single} as the dotted
line at the top of the figure.

In reality, stars have non-zero radii, and encounters of two stars with impact
parameter $Y\ij<Y\coll\equiv(R_i+R_j)[1+2G(m_i+m_j)/(R_i+R_j)w\ij^2]^{1/2}$
will lead to stellar collisions, where $R_i$ and $R_j$ are the radii of the
two stars $m_i$ and $m_j$.  Such collisions have two principal effects on the
calculation of the ejection rate: (i) Encounters with $Y\ij<Y\coll$ should not
be included in equation (\ref{eq:calR}). (ii) Stars cannot diffuse to energies
$E\la -Gm_*/R_*$ without colliding first, where $R_*$ is the stellar radius
\citep{BW76}. Thus,
we cut off the stellar distribution, so that $f(E)=0$ for $E<-Gm_*/R_*$. After
making these modifications, the ejection rate is hard to obtain analytically
so we resort to numerical calculations, shown by the dashed line in Figure
\ref{fig:single}.  As seen from Figure~\ref{fig:single}, the ejection rate is
significantly decreased by stellar collisions, especially at speeds
$\ga300\kms$ that approach or exceed the escape speeds from the stellar
surface.

These results do not include the effects of the gravitational potential of the
bulge, disk and halo. To include the effects of the bulge, we use equation
(\ref{eq:rho}) to compute the gravitational potential from the bulge; we add
the potential from the BH and use Eddington's formula to compute the bulge
DF $f(E)$; we then truncate the DF for
$E-\Phi_*(r=0)<-Gm_*/R_*$, and insert the result into the formula
(\ref{eq:calRs}) to obtain the ejection rate. The result is shown by the bold
solid line in Figure~\ref{fig:single}; this is much lower than the dashed line
at low speeds, because low-velocity stars cannot climb the potential to escape
the stellar bulge.  The difference between the bold solid line and the dashed
line is negligible at ejection speeds higher than $1000\kms$.

These results do not include the potential of the Galactic disk and halo.  The
gravitational potential difference between $R_0=8\kpc$ and the outer parts of
the bulge, $r\b\simeq 2\kpc$, can be roughly estimated by
\be
\Delta\Phi=V\c^2\ln(R_0/r\b)\simeq\half(370\kms)^2(V\c/220\kms)^2,
\label{eq:deltaPhi}
\ee
where $V\c$ is the circular speed, assumed independent of radius. The velocity
distribution of the stars that reach the solar radius $R_0$ is given by
\be
v\h=\sqrt{v_0^2-2\Delta\Phi}.
\label{eq:vh}
\ee
The thin solid line in Figure~\ref{fig:single} shows the ejection rate as a
function of $v\h$, which is obtained from the bold solid line using equations
(\ref{eq:deltaPhi}) and (\ref{eq:vh}). 

The ejection rate of stars at speeds $>100\kms$ is about (1--2)$\times
10^{-8}\yr^{-1}$ at the outer edge of the bulge and about $2\times
10^{-9}\yr^{-1}$ at the solar radius; the ejection rate of stars at speeds
$>500\kms$ is only $10^{-9}\yr^{-1}$ at the edge of the bulge and a factor of
three lower at the solar radius; and the ejection rate at speeds $>1000\kms$
is only $10^{-11}\yr^{-1}$ at either location. We conclude that the expected
rate of formation of hypervelocity stars in the Galaxy due to two-body close
encounters of single stars is negligible.

Finally, to illustrate the importance of the BH, we show the ejection rate
from a stellar system with the same density distribution as that in the
Galactic center but without a central BH (the dot-dashed line in
Figure~\ref{fig:single}). The existence of a central BH dramatically increases
the rate of ejecting hypervelocity stars.

\subsection{Tidal breakup of a binary star by the central BH}
\label{sec:binary}

\noindent
It is plausible that binary stars are formed as commonly near the Galactic
center as they are in the solar neighborhood.  Close encounters between a
close binary and a massive BH may break up the binary through an exchange
collision, in which one binary component becomes bound to the BH and the other
is ejected at high velocity (up to several thousand $\kms$;
\citealt{H88,GQ03}). 

We shall focus on binaries in the semimajor axis range $0.01\au<a\b<0.3\au$;
$0.01\au$ is nearly $2R_\odot$ and thus is a lower limit to the binary
separation for solar-type stars, and the choice of an upper limit of $0.3\au$
will be explained later. The binary period distribution observed locally by
\citet{dm91} implies that roughly 8\% of binaries belong to this period range;
given that the fraction of stars in binaries is $\sim 0.5$ we expect that the
binary fraction in the semimajor axis range $0.01\au<a\b<0.3\au$ is about
$0.04$. This estimate is quite uncertain, both because the estimate derived
for the solar neighborhood is uncertain and because we do not know whether the
binary distribution is the same at the Galactic center as it is in the solar
neighborhood.

We must first check that these binaries are not disrupted by encounters with
single stars. The disruption time is \citep{BT87}
\be
t_{\rm dis}=0.1{\sigma\over G\rho a\b\ln\Lambda},
\ee
where $\rho$ is the density of stars (eq.~\ref{eq:rho}), $a\b$ is the
semimajor axis of the binary, $\sigma$ is the velocity dispersion,
$\ln\Lambda$ is the Coulomb logarithm, and we have assumed that the binary
components and the field stars have the same mass. In the region $r\la 1\pc$
where the gravitational potential is dominated by the BH, we may take
$\sigma\simeq(GM\bh/r)^{1/2}$ and $\Lambda\simeq (M\bh/m_*)(a\b/r)$, so that
\be
t_{\rm dis}={4\times10^{10}\yr\over\ln[1.7(a\b/0.1\au)(1\pc/r)]}
\left(r\over1\pc\right)^{1.3}\left(0.1\au\over a\b\right).
\label{eq:tdis}
\ee
This formula is derived for soft binaries, for which $\Lambda\ga 1$; a similar
result holds for hard binaries, except that $\ln\Lambda$ is replaced by unity,
and $t_{\rm dis}$ represents the timescale in which the semimajor axis shrinks
by a factor of two due to encounters.  We conclude that binaries with
semimajor axis $a\b=0.3\au$ can survive for the age of the Galaxy outside
$r\simeq 0.8\pc$, while binaries with $a\b=0.01\au$ can survive outside $r\sim
0.05\pc$.

We now review Hills' (1988) three-body simulations of the encounter between a
binary and a massive BH. The center of mass of the binary is initially on an
unbound orbit relative to the BH, with velocity $v_\infty$. The probability of
an exchange collision is a decreasing function of the dimensionless
closest-approach parameter
\be
R_{\min}'\equiv \frac{R_{\min}}{a\b}\left(\frac{M\bh}{m_*}\right)^{-1/3};
\label{eq:Dmin}
\ee
here we have assumed for simplicity that $m_1=m_2=m_*$, and $R_{\min}$ is the
closest approach of the binary to the BH (the periapsis of the orbit of the
center of mass of the binary around the BH). The parameter $R_{\min}'$ is
essentially the distance of closest approach in units of the Roche radius of
the binary. Hills finds that exchange collisions occur in nearly 80\% of
encounters with $R_{\min}'\simeq 0.3$ and in 50\% of encounters with
$R_{\min}'\simeq 1$.  Let us define an ejection speed parameter by
(eqs.~\ref{eq:vdeltav} and \ref{eq:deltavb})
\be
v\bh'\equiv v\bh\left(\frac{m_*}{M\bh}\right)^{1/6}
\left(\frac{a\b}{0.1\au}\cdot\frac{\msun}{m_*}\right)^{1/2},
\label{eq:vej}
\ee
where $v\bh$ is the rms velocity at infinity of the ejected star. Then $v\bh'$
is mainly a function of $R_{\min}'$, and roughly independent of the BH mass
and the semimajor axis and masses of the binary components.  The velocity
$v\bh'$ varies between 130 and $160\kms$ for $R_{\min}'=0$--1, with a peak
near $R_{\min}'\simeq0.3$, and is about $130\kms$ at $R_{\min}'=1$
\citep{H88}.

These results are independent of the initial relative velocity $v_\infty$ of
the binary relative to the BH. This assumption is correct so long as this is
much less than the relative velocity at $R_{\min}$, which in turn requires
that\footnote{The dependence of the exchange cross-section on $v_\infty$ is
mapped out in more extensive three-body integrations by \citet{H91}, although
Hills' empirical summary of the results is based on the assumption that the
critical velocity scales as $M\bh^{1/6}$, not $M\bh^{1/3}$ as in equation
(\ref{eq:scale}).}
\be
v_\infty\la \left(Gm_*\over a\b\right)^{1/2}\left(M\bh\over
m_*\right)^{1/3}.
\label{eq:scale}
\ee
For the Galaxy this limit corresponds to $v_\infty\la
1.4\times10^4\kms(0.1\au/a\b)^{1/2}$, which is satisfied by a large margin.
Similarly, Hills' results for the probability of an exchange collision should
be valid for orbits that are bound to the BH so long as the orbits are
very eccentric, that is, $R_{\min}\ll a$ where $a$ is the semimajor axis of the
orbit around the BH. 

For numerical estimates we assume that both binary components have the solar
mass. According to equation (\ref{eq:Dmin}), for a binary to have an exchange
probability of 50\% or more in an encounter with a $3.5\times 10^6\msun$ BH,
it must pass within about $R_{\min}\simeq 150a\b$ of the BH; that is, within
$R_{\min}=1.5$ or $45\au$ for $a\b=0.01$ and $0.3\au$ respectively.  The rms
ejection speed at infinity $v\bh$ in this case is about $5\times10^3$ or
$0.9\times 10^3\kms$ for $a\b$=0.01 or $0.3\au$ respectively.  These ejection
speeds for $a\b\la0.3\au$ are high enough to escape the Galactic bulge and
even the Galactic halo, which is why we restrict our attention to this range
of binary semimajor axis. 

The ejection rate depends on the flux of binary stars passing close to the
central BH, specifically, within a periapsis distance $R_{\min}\simeq 150a\b$.
The region in the $(J,E)=$(specific angular momentum, specific energy) phase
space where a binary star has dimensionless closest approach parameter
$R_{\min}'\la 1$ is called the ``loss cone'' and is given by:
\be
J^2\le J\lc^2(E,a\b)\equiv 2 R_{\min}^2\left[E-\Phi(R_{\min})\right]
\simeq 2GM\bh R_{\min} \qquad (|E|\ll GM\bh/R_{\min}).
\label{eq:losscone}
\ee
[Note that the concept of the ``loss cone'' is used in other contexts, for
example, see \citet{MT99} for the tidal disruption of single stars by a
massive BH or \citet{Y02} and \S\ref{sec:bbh} for the expulsion of
low-angular momentum stars by a BBH.  The loss cone for disruption of single
stars is usually much smaller than the loss cone for disruption of binary
stars.]  At first, the ejection rate is determined by the rate of depletion of
the initial population of binary stars in the loss cone, and the rate per unit
energy is given by (see eq.~19 in \citealt{Y02}):
\be
F\full(E,a\b)\simeq 4\pi^2\eta f(E)J\lc^2(E,a\b),
\qquad t\la P(E),
\label{eq:Ffull}
\ee
where $P(E)$ is the radial period of an orbit with energy $E$ and zero angular
momentum, and $\eta$ is the number fraction of binary stars.  As the binary
stars in the loss cone are disrupted, new binaries are scattered into the loss
cone by two-body relaxation. Eventually, the scattering into the loss cone
reaches a steady state.  The steady-state diffusion rate of binary stars into
the loss cone per unit energy is obtained from the 
Fokker-Planck equation (eq.~24 in \citealt{MT99}):
\be
F\lc(E)\simeq
\cases{ q F\full(E,a\b)/\ln[GM\bh/(4|E|R_{\min})],
& if $q\ll -\ln(J\lc^2/J\c^2)$\cr
F\full(E,a\b), & if $q\gg -\ln(J\lc^2/J\c^2)$},
\label{eq:Fmax}
\ee
where 
\be
q\equiv P(E)\bar{\mu}(E)J^2\c(E)/J^2\lc(E,a\b).
\label{eq:q}
\ee
Here $\bar{\mu}$ is the orbit-averaged diffusion coefficient and $J^2\c(E)$ is
the specific angular momentum of a circular orbit at energy $E$. The quantity
$\half qJ^2\lc$ is the mean-square change in the scalar angular momentum for a
nearly radial orbit in one orbital period; $\half qJ^2\lc=P(E)\langle (\Delta
J)^2\rangle=P(E)\langle r^2(\Delta v_{\rm t})^2\rangle$, where
$\langle\cdot\rangle$ denotes an orbit-averaged diffusion coefficient and
$v_{\rm t}$ is the tangential velocity (see details in Appendix B of
\citealt{MT99}). The region in which $q\ga
-\ln(J\lc^2/J\c^2)$ is known as the ``pinhole'' or ``full loss cone'' regime,
as opposed to the ``diffusion'' or ``empty loss cone'' regime. Generally the
loss cone is full at large radii and empty at small radii, we shall define the
critical radius $r_{\rm crit}$ to be the apoapsis of orbits at the transition
between these two regimes. The critical radius $r_{\rm crit}\sim 600\pc$
for $a\b=0.3\au$ and $\sim 20\pc$ for $a\b=0.01\au$.

We can use the model for the Galactic center described in
\S\ref{sec:Milky} to estimate the ejection rate due to disruption of binary
stars. As usual we assume that the DF is isotropic in velocity space, and that
it vanishes for energy $E-\Phi_*(r=0)\simeq E <-G\msun/\rsun$ (without this
cutoff the 
ejection rate for a full loss cone diverges). If the binary loss cone is full,
the ejection rate is given by equation (\ref{eq:Ffull}):
\begin{eqnarray}
n\ej&=&\int F\full(E)\,\d E\nonumber \\
{}&\simeq& 8\pi^2 \eta G M\bh R_{\min} C\int_{-G\msun/\rsun}^0|E|^{\alpha-3/2}\d E\nonumber \\
{}&\simeq & 2\times 10^{-2}\yr^{-1} \left(\eta\over
0.1\right)\left(a\b\over 0.1\au\right)\left( R'_{\rm min}\over 1\right)
\left(\frac{M\bh}{3.5\times10^6\msun}\right)^{4/3-\alpha},
\end{eqnarray}
which is approximately consistent with the estimate $\sim
10^{-2}(\eta/0.1)(a\b/0.1\au)$ in \citet{H88}, based on much less certain
observational parameters. However, this calculation, like Hills', neglects the
fact that the binaries initially in the loss cone are rapidly depleted. For a
more accurate ejection rate we must use equation (\ref{eq:Fmax}), which yields
\be
n\ej=\int F\lc(E)\,\d E\simeq 1.5\times 10^{-5}(\eta/0.1)\yr^{-1}
\label{eq:nej}
\ee
if $a\b=0.3\au$. The ejection rate is quite insensitive to the binary
semimajor axis $a\b$, decreasing to only $n\ej=0.9\times
10^{-5}(\eta/0.1)\yr^{-1}$ if $a\b=0.01\au$.
The apoapsides of most of the stars contributing to the total
diffusion rate into the loss cone are in the range 2--3$\pc$ for
$a\b=0.01$--$0.3\au$; at these radii the binary disruption rate due to
encounters with field stars (eq.\ \ref{eq:tdis}) is small. Because the
loss cone is not full, Hills' simple estimate of the ejection rate is too
large by three orders of magnitude at $a\b=0.1\au$.

The time required for the ejected stars to travel a distance $D$ is
by: 
\begin{eqnarray}
t\D & \simeq & \frac{D}{v\bh}=5\times 10^6\yr
\left(\frac{D}{8\kpc}\cdot\frac{1.6\times10^3\kms}{v\bh}\right)\label{eq:tD}\\
& \sim & 5\times10^6\yr
\left(\frac{D}{8\kpc}\cdot\frac{1.3\times 10^2\kms}{v\bh'}\right)
\nonumber \\
& & \times\left(\frac{3.5\times10^6\msun}{M\bh}
\cdot\frac{m_*}{\msun}\right)^{1/6}
\left(\frac{a\b}{0.1\au}\cdot\frac{\msun}{m_*}\right)^{1/2},
\label{eq:tDb}
\end{eqnarray}
where $v\bh'$ is the ejection speed parameter defined in equation
(\ref{eq:vej}). From equations (\ref{eq:nej}) and (\ref{eq:tDb}), the number
of hypervelocity stars inside the sphere with radius $D$ is given by:
\be
n\ej t\D\sim 60\left(\frac{\eta}{0.1}\right)\left(\frac{D}{8\kpc}\right)
\left(\frac{a\b}{0.1\au}\right)^{1/2}
\left(\frac{\msun}{m_*}\right)^{1/3}.
\label{eq:nejHills}
\ee

In this process, one component of the binary star is ejected with an energy
gain, and the other loses energy and becomes more tightly bound to the BH.
The apoapsis distance of the bound star is approximately $GM\bh/v\bh^2 \simeq
1\times 10^3\au (M\bh/3.5\times 10^6\msun)(1.6\times10^3\kms/v\bh)^2$.  The
eccentricity of the bound star is usually high, and the corresponding orbital
period is $\sim 8\yr(M\bh/3.5\times
10^6\msun)(1.6\times10^3\kms/v\bh)^3$. Recent high-resolution infrared
observations of the region around the compact radio source (and presumed BH)
$\sgr$ have revealed massive young stars on high-eccentricity orbits with
periods of decades or less
\citep{Schodel02,Schodel03,eis03,Ghez03a,Ghez03b,Ghez03c}. \citet{GQ03} have
suggested that these unusual orbits may have been produced by Hills'
mechanism; that is, these stars may be one component of a tidally disrupted
binary star.  According to equation (\ref{eq:nej}), we expect that the number
of young ($10^7\yr$ old) stars that are the remnants of tidally disrupted
binaries should be $\sim 100(\eta/0.1)$ at any time. Of course, the number of
high-mass remnants would be smaller, by an amount that depends on the shape of
the initial mass function.

\subsection{Interactions between a single star and a massive BBH}
\label{sec:bbh}

\noindent
BBHs are likely to exist in some galactic centers \citep{BBR80,Y02}, and
perhaps in the Galactic center as well. In fact, 
\citet{hm03} have recently argued that the young stars observed at 0.1--1
$\mpc$ distances from $\sgr$ may have been shorn from a disk or cluster
surrounding the smaller component of a BBH, by tidal forces from the larger
component.  A massive BBH steadily loses orbital energy to the surrounding
stellar population.  When the two BHs are orbiting independently in the
galaxy, and even after they form a bound binary system, the energy loss is
mainly through dynamical friction.  Once the BBH becomes sufficiently tightly
bound that its orbital velocity exceeds the velocity dispersion of the stars,
dynamical friction becomes less and less effective, and three-body
interactions with low-angular momentum stars passing through the BBH become
the dominant energy-loss mechanism. When the BBH becomes
``hard'', at a semimajor axis \citep{Q96}
\be
a\h\equiv \frac{GM_2}{4\sigma\c^2}\simeq 0.04\pc\left(\frac{\nu}{0.1}\right)
\left(\frac{M_1+M_2}{3.5\times10^6\msun}\right)
\left(\frac{100\kms}{\sigma\c}\right)^2,
\label{eq:ah}
\ee
where 
\be
\nu\equiv M_2/(M_1+M_2)
\label{eq:nu}
\ee
is the mass ratio of the small BH to the BBH, most of the low-angular
momentum stars that pass near a BBH will eventually be expelled with an energy
gain after one or several encounters with the BBH. In such expulsions, the
average ejection speed of the stars is
\be
v\bh\simeq \sqrt{\frac{3.2GM_1M_2}{(M_1+M_2)a\bh}}
\simeq 2.2\times10^{3}\kms\left(\frac{\nu}{0.1}\right)^{1/2}
\left(\frac{1\mpc}{a\bh}\right)^{1/2}
\left(\frac{M_1}{3.5\times10^6\msun}\right)^{1/2}
\label{eq:vbbh}
\ee
(see eq.~17 in \citealt{Y02}). The average relative energy change $\Delta
{\cal E}/{\cal E}$ per ejection is independent of the BBH orbital energy
${\cal E}$.  As the semimajor axis of the BBH continues to decrease,
eventually gravitational radiation takes over as the dominant energy loss
mechanism.

We now ask what are the observational constraints on the mass ratio $\nu$ and
semimajor axis $a\bh$ of a possible BBH in the Galactic center. We shall
assume that the total mass $M_1+M_2=3.5\times10^6M_\odot$, that the
BBH is on a circular orbit, and that $\sgr$ is the larger component $M_1$
of the BBH.  (We shall argue below that $\sgr$ is unlikely to be the small
component of the BBH.) For reference, the orbital period of the BBH is
\be
P\BBH=\frac{2\pi a\bh^{3/2}}{\sqrt{G(M_1+M_2)}}
\simeq 1.6\yr\left(\frac{a\bh}{1\mpc}\right)^{3/2}
\left(\frac{3.5\times 10^6\msun}{M_1+M_2}\right)^{1/2}
\label{eq:PBBH}
\ee

There are several constraints on the properties of the BBH:

\begin{itemize}

\item A census of stars within $\sim 10\arcsec\simeq0.4\pc$ of the Galactic
center reveals that the peak of the stellar surface density agrees with the
position of $\sgr$ within $\sim 0\,\farcs2\simeq 8\mpc$
\citep{Genzel03}. Taking the center of mass of the BBH as the center of the
stellar distribution, and ignoring the small possibility that the two BHs are
aligned along the line of sight to us, we have the following constraint on the
distance between $\sgr$ and the center of mass of the BBH:
\be
\nu a\bh\la 0\,\farcs2\simeq 8\mpc.
\label{eq:center}
\ee
In panel (b) of Figure~\ref{fig:bbh}, the region below the bold long-dashed
line gives the parameter space represented by inequality (\ref{eq:center}).

\item The eccentric Keplerian orbits of stars at radii $r\sim
0.1$--$1\mpc$ 
\citep{Schodel02,Schodel03,Ghez03a,Ghez03b,Ghez03c} will eventually provide 
strong constraints on the properties of a BBH at the Galactic center. However,
a binary with semimajor axis small compared to the periapsis distance of the
stellar orbit, or large compared to the apoapsis distance, will be poorly
constrained because the orbit will be nearly a Keplerian ellipse, with one
focus at either the center of mass or one component of the BBH,
respectively. Moreover, so far these stars have only been observed for a
fraction of an orbital period, so the sensitivity to deviations from a
Keplerian orbit is still relatively small (but growing rapidly). Without
carrying out proper simulations, we guess that the agreement of the
observations to date with orbits around a single point mass constrains the
mass fraction to be $\nu\la 0.2$ if $a\bh$ is in the range $\sim
0.1$--$1\mpc$. According to this constraint, the BBH is unlikely to be in the
box enclosed by the dot-long-dashed line near the right boundary of panel (b)
of Figure~\ref{fig:bbh}.

\item Observations of the proper motion of $\sgr$ also constrain the
parameter space $(a\bh,\nu)$ of a putative BBH in the Galactic center.  VLBA
observations from 1995 to 2000 show that the dominant term in the proper
motion of $\sgr$ with respect to extragalactic radio sources comes from the
motion of the Sun around the Galactic center. The rms residual from uniform
motion is $\Delta\simeq0.5\mas=0.02\mpc$, and the upper limit of the
peculiar motion of $\sgr$ perpendicular to the Galactic plane is
$v_\perp\simeq 8\kms$ \citep{BS99,Reid99,Reid03}. We do not know the motion of
the Sun in the Galactic plane with comparable accuracy, but unless the
peculiar velocity of $\sgr$ happens to be closely aligned with the Galactic
plane, it is unlikely to be larger than $\sqrt{2}v_\perp\simeq v_\max
\simeq 12\kms$. If the orbital period of the BBH $P\BBH$
(eq. \ref{eq:PBBH}) is short compared to the observation interval $T=5\yr$,
then the amplitude of the motion of $\sgr$ relative to the center of mass of
the BBH must be smaller than the rms observational error, that is,
\be
\nu a\bh\la f\Delta,
\label{eq:Delta}
\ee
where $f$ is of order unity. In the opposite limit, if
$P\BBH \gg T$ then the velocity 
of $\sgr$ relative to the center of mass must be less than $v_\max$, so
\be
\nu a\bh\la f v_\max P\BBH/(2\pi).
\label{eq:vpec}
\ee
We set $f\simeq 2$ and simply choose the transition between inequalities
(\ref{eq:Delta}) and (\ref{eq:vpec}) so that they are continuous.
In panel (b) of Figure~\ref{fig:bbh}, the region on the left side of the bold
solid line gives the parameter space allowed by inequalities (\ref{eq:Delta})
and (\ref{eq:vpec}) with $\Delta=0.5\mas$ and $v_\max= 12\kms$
\citep{Reid03}. 

\end{itemize}

The rate of ejecting stars by interactions with a hard BBH depends on the
flux of low angular momentum stars passing through the vicinity of the BBH.
Just as in the case of the loss cone of binary stars discussed in
\S\ref{sec:binary}, here the region in the (specific energy, specific angular
momentum) phase space in which a star can pass within $\sim a\bh$ of the BBH
is called the loss cone, and may be found by replacing $R_{\min}$ with the
semimajor axis of the BBH in equation (\ref{eq:losscone}) (see \citealt{Y02}).
Initially, the loss cone of the BBH is full.  As stars are removed from the
loss cone by close encounters with the BBH, new stars refill the loss cone by
two-body relaxation (or by non-axisymmetric gravitational forces from the
surrounding stellar system, if it is non-spherical; 
see \citealt{Y02}).  Eventually the system reaches a steady
state controlled by the balance between the ejection rate of stars and the
rate at which stars refill the loss cone.  If the rms angular momentum
transfered to or from the stars per orbital period is larger than $J\lc$ (see
eq.~\ref{eq:losscone}), then the stars will refill the loss cone as fast as it
is depleted and the loss cone remains full; otherwise, the stars will slowly
diffuse into the loss cone and the loss cone remains nearly empty.  The
refilling rate due to two-body relaxation can be found by solving the
steady-state Fokker-Planck equation (e.g.\ eq.~17 in
\citealt{MT99} or see \citealt{Y02}).  If the BBH is hard and has a total mass
$3.5\times 10^6\msun$, our calculations show that the rate of removing stars
from the loss cone is $n\full\BBH\simeq 2\yr^{-1}(a\bh/1\mpc)$ when the loss
cone is full and $n\diff\BBH\simeq$(1--3)$\times 10^{-4}\yr^{-1}$ when the
loss cone is empty; the superscript ``diff'' refers to the fact that the
refilling of the loss cone is a diffusive process. The rate $n\full\BBH$ is
insensitive to the BBH mass ratio $\nu$, and the rate $n\diff\BBH$ is
insensitive to both $a\bh$ (only through the logarithmic term in
eq.~\ref{eq:Fmax}) 
and $\nu$, so long as the BBH is hard and the mass
of the smaller component of the BBH is significantly larger than the stellar
mass (Quinlan's 1996 simulations only cover the range $\nu>0.004$ but his
finding that the hardening is largely independent of $\nu$ should extend to
more extreme mass ratios).  The hardening timescale of the BBH is then given
by (see eq.~\ref{eq:vbbh}):
\be
t\h(a\bh)=\left|\frac{a\bh}{\dot a\bh}\right|\simeq\frac{M_1+M_2}
{3m_*n\diff\BBH}
\simeq 6\times 10^9\yr\left(\frac{M_1+M_2}{3.5\times 10^6\msun}\right)
\left(\frac{1\msun}{m_*}\right)
\left(\frac{2\times 10^{-4}\yr^{-1}}{n\BBH\diff}\right),
\label{eq:th}
\ee
which is shown as a function of $a\bh$ and $\nu$ in Figure
\ref{fig:bbh} (see the solid horizontal lines in panels c and d, which
represent 4, 6, $8\times10^9\yr$ from top to bottom). Assuming that the BBH
orbit is circular, its gravitational radiation timescale is given by
\citep{P64}:
\be
t_{\rm gr}(a\bh,\nu)=\left|\frac{a\bh}{\dot a\bh}\right|=\frac{5}{64}
\frac{c^5a\bh^4}{G^3M_1M_2(M_1+M_2)}
\simeq 5.4\times 10^8\yr\left(\frac{a\bh}{1\mpc}\right)^4
\frac{(3.5\times10^6\msun)^3}{(M_1+M_2)^2M_1}
\left(\frac{0.1}{\nu}\right),
\label{eq:tgr}
\ee
which is also shown in panels (c) and (d) of Figure \ref{fig:bbh} (the
long-dashed lines, which represent $10^9,10^8,10^7\yr$ from top to bottom). A
similar diagram to panel (c) is shown in \citet{hm03}. The BBH loses energy
mainly by interacting with stars when $t_{\rm gr}(a\bh,\nu)>t\h(a\bh)$ and
mainly by gravitational radiation when $t_{\rm gr}(a\bh,\nu)<t\h(a\bh)$.

If we do not live at a special time, the lifetime of the BBH orbit,
$\min(t\h,t_{\rm gr})$, should exceed $\nu\times 10^{10}\yr$ (i.e. at most,
the central BH should accrete a fraction $\nu$ of its mass in a fraction $\nu$
of its age). This constraint is shown as the bold
long-dashed line in panels (c) and (d) of Figure \ref{fig:bbh}. We may combine
this with the constraints from the position of $\sgr$ (inequality
\ref{eq:center}), the near-Keplerian orbits of stars close to $\sgr$, and the
proper motion of $\sgr$ (inequalities~\ref{eq:Delta} \ref{eq:vpec}),
to conclude that the BBH should not be located in the dotted
region in panel (d) of Figure~\ref{fig:bbh}.  Note that we have assumed that
$M_1$ coincides with $\sgr$ and is the larger component of the BBH, so that
$\nu$ cannot exceed 0.5 (eq.\ \ref{eq:nu}). However, if we relax this
constraint, that is, if $\nu$ is allowed to exceed 0.5, we may extend the
above constraints on the BBH parameter space and find that the forbidden
region in Figure~\ref{fig:bbh}(d) excludes the range $0.5<\nu<1$. Thus, $\sgr$
cannot be the smaller component of the BBH.

The bold dot-short-dashed line in Figure~\ref{fig:bbh} (panels a and
d) represents the semimajor axis at which the BBH becomes hard, given
by equation (\ref{eq:ah}). The average ejection velocity $v\bh$ as a
function of $a\bh$ and $\nu$ is shown by short-dashed lines in these
panels (these represent $v\bh=400$--$2000\kms$ with interval $\delta
v\bh=200\kms$ from top to bottom, with $v\bh=1000\kms$ shown in bold).
Some short-dashed lines with $v\bh>1000\kms$ cross the allowed region
of panel (d), and hence there are allowed parameters for a BBH that
could eject hypervelocity stars that reach the solar radius. 

Thus, for example, consider a BBH with $\nu=0.01$ and
$a\bh=0.5\mpc$. This has orbital period $P\BBH\simeq 0.6\yr$,
gravitational radiation timescale $t_{\rm gr}\simeq 4\times10^8\yr$,
and distance of $\sgr$ from the center of mass of 0.1 mas, consistent
with the observational constraints we have described.  The time for a
hypervelocity star to travel $D=8\kpc$ is $t_D\simeq 8\times
10^6\yr(D/8\kpc)(10^3\kms/v\bh)$ (eq.~\ref{eq:tD}). Since
$t\D\ll\min(t\h,t_{\rm gr})$, we can ignore the BBH orbital evolution,
so the number of stars with ejection speed higher than $10^3\kms$
within the sphere $D<8\kpc$ can be roughly estimated to be $n\diff\BBH
t\D\simeq 10^{3}$. 

\begin{figure}
\epsscale{0.8}
\plotone{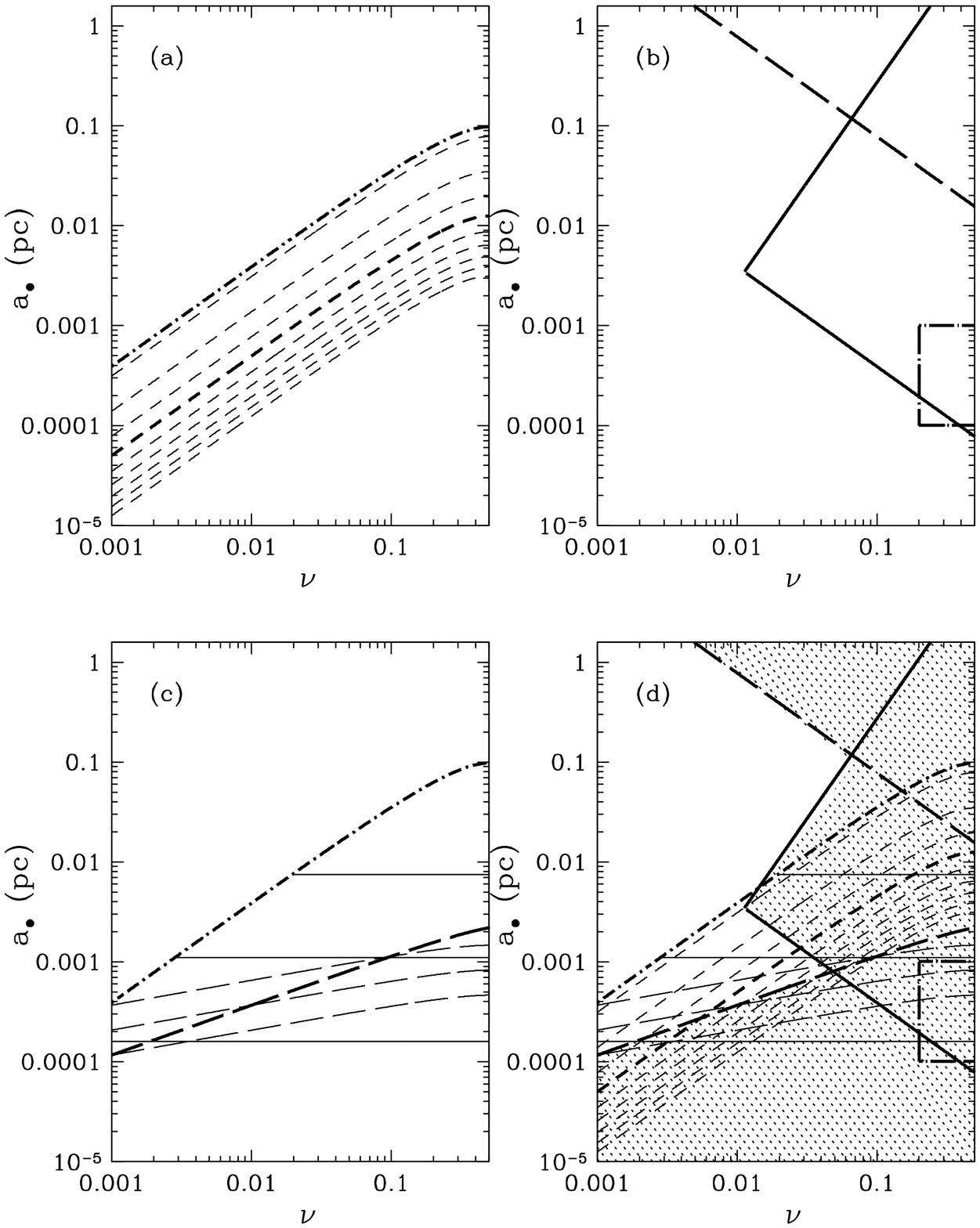}
\end{figure}
\begin{figure}
\caption{In each of the four panels, the horizontal axis represents
the mass ratio $\nu\equiv M_2/(M_1+M_2)$ of a hypothetical BBH in the Galactic
center and the vertical axis represents its semimajor axis $a\bh$.  $\sgr$ is
taken to be the more massive component of the BBH.  The eccentricity of the
BBH orbit is assumed to be zero.  Panel (a): the bold dot-short-dashed line
gives the semimajor axis of the BBH when it becomes hard (see eq.~\ref{eq:ah},
with $\sigma\c=100\kms$).  The short-dashed lines represent the average
ejection speeds of the stars from the BBH (eq.~\ref{eq:vbbh}),
$v\bh=400$--$2000\kms$ from top to bottom with interval $200\kms$, with
$v\bh=1000\kms$ in bold.  Panel (b): the region above the bold long-dashed
line is the BBH parameter space that is excluded by the location of $\sgr$
close to the center of the stellar cusp (see inequality~\ref{eq:center}).  The
rectangular region enclosed by the bold dot-long-dashed line is excluded by the Keplerian
orbits of stars at radii $r\sim 0.1$--$1\mpc$.  The region on the right side
of the bold solid line is excluded by the proper motion of $\sgr$, except for
unusual orientations in which its motion is almost along the line of sight
(inequalities~\ref{eq:Delta} and \ref{eq:vpec}).  The kink at $a\bh\simeq 5\mpc$
represents an approximate boundary between the cases in which the period of
the BBH is long or short compared to the present duration of proper-motion
observations. Panel (c): the long-dashed lines give the gravitational
radiation timescale of the BBH: $10^9,10^8,10^7\yr$ from top to bottom
(eq.~\ref{eq:tgr}). The solid lines give the hardening time $t\h$: 4, 6,
$8\times10^9\yr$ from top to bottom (eq.\ \ref{eq:th}). The region above the
bold long-dashed line represents the constraint that $\min(t\h,t_{\rm gr})>
\nu\times 10^{10}\yr$, so that the larger BH accretes at most a fraction $\nu$
of its mass in a fraction $\nu$ of its age (i.e. the present accretion rate of
BH companions is not unusually large compared to the average over the lifetime
of the Galaxy).  As in panel (a), the bold dot-short-dashed line gives the
semimajor axis of the BBH when it becomes hard. Panel (d): combination of the
curves in panels (a)--(c).  The shaded region is excluded.  See
\S\ref{sec:bbh}.}
\label{fig:bbh}
\end{figure}

\section{Discussion and conclusions}\label{sec:discon}

\noindent
We have studied stellar dynamical processes that eject hypervelocity
($>10^3\kms$) stars from the Galactic center.  We consider three
mechanisms of ejecting stars: close encounters of two single stars, tidal
breakup of binary stars, and three-body interactions between a star and a BBH.

The rate of ejection from close encounters of single stars is strongly
affected by the finite size of stars: Figure \ref{fig:single} shows
that this rate is $10^5$ times smaller for solar-type stars than for
point masses with the same radial distribution. For solar-type stars,
the rate of ejection with speed $\ga 10^3\kms$ at $R_0=8\kpc$ is only
$10^{-11}\yr^{-1}$, which is too low to be detectable. If there is a
distribution of stellar masses, the ejection rate is likely to be larger for
the lighter stars \citep{H69}, but these also have lower luminosity and thus
are harder to detect. 

The rate of ejection from breakup of binary stars \citep{H88} depends on the
orbital parameters of the binaries, such as the semimajor axis and component
masses.  We estimate that the rate of ejecting hypervelocity stars is
$10^{-5}(\eta/0.1)\yr^{-1}$ (eq.~\ref{eq:nej}), where $\eta$ is the binary
fraction and the result is only weakly dependent on semimajor axis
for $a\b\la 0.3\au$. This rate is also
not strongly dependent on the stellar mass distribution.  With the assumption
that the binary components have the same mass $m_*$, the number of
hypervelocity stars within the solar radius is $\sim 100$ for $\eta=0.1$ and
$a\b=0.1\au$ (eq.~\ref{eq:nejHills}).  

$\sgr$ may be one component of a BBH \citep{hm03}. We discuss the constraints
on the parameter space (mass ratio and semimajor axis) of the BBH using a
variety of observational and theoretical arguments.  The ejection speed of
stars due to three-body interactions with the BBH depends on the BBH semimajor
axis and mass ratio, but not on stellar mass.  In part of the allowed
parameter space, near $a\bh=0.5\mpc$ and $\nu=0.01$, the average
stellar ejection speed can exceed $10^3\kms$ (see panel (d) of
Fig.~\ref{fig:bbh}), and the rate of ejecting hypervelocity stars is $\sim
10^{-4}\yr^{-1}$. In this case the expected number of stars with ejection
speeds higher than $10^3\kms$ within the solar radius $r<8\kpc$ is $\sim
10^3$. 

Nearby high-velocity halo stars bound to the Galaxy cannot have velocities
that exceed the local escape speed. Recent estimates of the escape speed are
$550\pm100\kms$ \citep{lt90}, $610\pm120\kms$ \citep{k96}, and $600\kms$
\citep{we99}, well below our cutoff at $10^3\kms$, so it is unlikely that
bound halo stars can appear to be hypervelocity stars. Rare Local Group
interlopers, perhaps from M31, could have velocities exceeding the escape
speed. However, hypervelocity stars ejected from the Galactic center have an
additional distinctive kinematic signature, since their velocity vectors
should point almost directly away from the Galactic center (apart from small
changes induced by the non-spherical component of the Galactic potential).

Hypervelocity stars produced at the Galactic center should also be
distinguishable from runaway stars produced by supernova explosions in close
binary systems or by close encounters in star clusters \citep{leo93}, which
will have quite different velocity distributions. 

At a distance of $10\kpc$, a solar-type star traveling at $10^3\kms$ has
apparent magnitude $V=19.8$ and proper motion $20\mas\yr^{-1}$. Ground-based
surveys can achieve proper-motion accuracies of $\sim 1\mas\yr^{-1}$ with
multiple images over baselines of a few years \citep{ala03}. Planned
ground-based surveys such as Pan-STARRS and the Large Synoptic Survey
Telescope, with limiting magnitude $V\simeq 24$, could easily detect and
measure the proper motion of solar-type hypervelocity stars over much of the
Galaxy. The GAIA spacecraft will survey the whole sky to $V=20$ with
proper-motion accuracy of $\sim 0.1\mas\yr^{-1}$ or better at its limiting
magnitude.

In Hills' mechanism, one component of the binary star is ejected, and the
other is thrown onto a high-eccentricity orbit around the central BH
\citep{GQ03}.  We expect that the number of young ($10^7\yr$ old) stars that
are the remnants of tidally disrupted binaries should be $\sim 100 (\eta/0.1)$
at any time.

Hypervelocity stars produced by Hills' mechanism should be approximately
spherically distributed, while the spatial distribution of the hypervelocity
stars due to interactions with the BBH should be flattened in the orbital
plane of the BBH (see \citealt{ZB01}).

Support for this research was provided in part by NASA through grants from
the Space Telescope Science Institute, which is operated by the Association of
Universities for Research in Astronomy under NASA contract NAS5-26555.

\end{document}